# Effect of oxygen adsorption and oxidation on the strain state of Pd nanocrystals

Binayak Mukherjee, Alberto Flor & Paolo Scardi

*Department of Civil, Environmental and Mechanical Engineering, University of Trento, Via Mesiano 77, 38123 Trento, Italy*

**Abstract**

X-ray powder diffraction using a synchrotron light source reveals significant modifications to both morphology and strain state in Palladium nanocubes after oxidation. Short-range strain measured by the static component of the Debye-Waller coefficient is observed to be higher in the oxidized nanoparticles; while long-range strain related to the line broadening of the diffraction peaks is seen to decrease. Using multiscale modelling with classical molecular dynamics and density functional theory, we connect the decrease in long-range strain to the increased truncation of the oxidized nanocubes, while the higher short-range strain is shown to be due to surface softening from oxygen adsorption. Different surface disorder on different crystallographic facets lead to opposing trends for oxygen activation on the different exposed surfaces of the truncated nanoparticles.

## 1. Introduction

Palladium nanocrystals (Pd NC's) are well known as catalysts in the chemical industry, with a wide range of applications including oxidation of organic molecules[1], synthesis of polymeric materials[2,3], hydrogenation reactions[4,5], and the oxygen reduction reaction in fuel cells[6]. Among organic molecules, the oxidation of methane ($CH_4$) to carbon dioxide ($CO_2$) is considered an environmentally sustainable combustion process compared with other hydrocarbon based fuels, due to the comparatively lower carbon-to-hydrogen ratio in $CH_4$. However, among all saturated hydrocarbons, methane has the highest carbon-hydrogen bond strength which in turn produces sub-optimal ignition performance for reaction in the gas phase[7]. On the other hand, $CH_4$ dissociates with low activation energy on several noble metal surfaces, allowing an enhancement of the ignition reaction. Of these noble metals, Pd is regarded as one of the most promising catalysts for complete, low temperature oxidation of $CH_4$ [8–11]. The catalytic activity of Pd nanocrystals is known to be a function of size, morphology, synthesis method and oxidation state of the particle, and as such, these factors play an important role in determining the efficiency of the $CH_4$ oxidation process. Additionally, due to the exposure to ambient oxygen ($O_2$) throughout the combustion process, Pd-based catalysts are expected to be oxidized and form both surface and possibly some bulk oxide, a phenomenon which has been shown to measurably affect the catalytic process[12–14]. Reports in the literature suggest that PdO is the more catalytically active phase[15] at lower temperatures, while at higher temperatures metallic Pd dominates catalytic activity[16]. Compared to pure metallic Pd, coincident Pd and PdO phases have been shown to be more active catalysts for the combustion reaction of $CH_4$ [17].

Another industrially important catalytic process which involves the adsorption of $O_2$ on Pd surfaces is the oxygen reduction reaction (ORR). Pd has been used as the cathode material for ORR in electrochemical applications such as fuel cells. ORR has been studied in alkaline media using anion exchange membranes (AEMs). A significant motivation to change the electrolyte membrane from acidic (mainly in proton exchange membranes (PEMs)) to alkaline is the improved electrokinetics of ORR in an alkaline environment, in contrast to the excessive corrosion in acidic media[18,19]. In this regard, several cathode electrocatalysts have been proposed, among which Pd-based NC's are an important candidate [20].

Several reports in the literature have addressed the adsorption of oxygen and/or subsequent oxidation of the metal surface of Pd NC's. Collins et al (2014)[21] have studied the oxidation of Pd NC's using a combination of transmission electron microscopy (TEM) and x-ray photoelectron spectroscopy (XPS). They have demonstrated that oxidation of the NC starts from edge and corner sites, leading cubic NC's to evolve towards a rounded shape, whereas octahedral particles tend to maintain their morphology. Using density functional theory (DFT) calculations alongside high performance liquid crystallography (HPCL) and near-edge X-ray absorption fine structure (NEXAFS) experiments, Long et al (2013)[22] have shown that the activation of $O_2$ for catalytic processes by Pd NC's is strongly dependent on the corresponding crystallographic facet. They have identified the 100 surface of the NC to be the most efficient in inducing a spin-flip process in the $O_2$ molecule via a transfer of charge from the Pd surface to the molecule, thereby activating the inert ground-state $O_2$ molecule to a highly reactive singlet $O_2$. Using molecular dynamics (MD) simulations with a reactive force-field (ReaxFF), Mao et al (2016)[23] have studied the role of Pd NC's in the combustion of methane. The authors have predicted that $O_2$ molecules block the active sites for the adsorption of $CH_4$, and are less likely to dissociate on oxygen-coated Pd surfaces. Significantly however, these same oxygen-coated NC surfaces induce the dissociation of $CH_4$ at a lower temperature compared to bare NC's. Most recently, Zhang et al (2017) [24] have confirmed using high-resolution TEM studies that oxidation of Pd NC's starts from active sites present at the edge and corners of the particle, and have shown that the oxide phase starts by forming small, island-like clusters on the NC surface.

While these aforementioned studies and others have thoroughly characterized the kinetics of oxygen adsorption on the surface of Pd NC's, they remain conspicuously silent on the matter of strain generated due to the presence of oxygen on the NC surface. Surface lattice strain is well known for being a crucial factor which may be tuned in order to enhance catalytic performance in NC's[25]. Lattice distortion due to surface strain is known to modify the electronic structure of the metal NC surface by inducing a shift in the surface d-band center[25]. As a result, strain-engineering of the NC surface can be effectively used to optimize the sorption energies of molecules for a particular chemical reaction[25]. Combining coherent diffraction imaging (CDI) with reactive MD and finite elements modelling, Kim et al (2018)[26] have studied the role of oxygen adsorption in modifying the strain-state of Pt nanocatalysts during $CH_4$ oxidation. In a recent publication, Rebuffi et al (2020)[27], through a combination of X-ray powder diffraction (XRPD) experiments and MD and DFT simulations, have studied the role of capping agent-induced disorder on a Pd NC surface. They have conclusively shown that the short-range strain developed on the surface of a Pd NC due to surface softening induced by the adsorption of the capping agent CTAB is significant enough to overcome the strain-effects associated with both size and shape of the nanoparticle. As such, the question of how the strain-state of a Pd NC is affected by the presence of oxygen is highly relevant for understanding the effects of oxygen adsorption and oxidation on the catalytic performance of Pd NC's.

In this article, we aim to answer this question through a combination of experimental results from XRPD, and theoretical modelling in the form of multiscale simulations using MD with reactive potentials, as well as DFT. The tried-and-tested XRPD technique, with high quality data obtained from synchrotron light sources and analysed using advanced data analysis paradigms such as the whole powder pattern modelling (WPPM) approach[28], can prove a unique asset in characterizing strain in a NC. By completely and accurately fitting high quality diffraction data, the short- and long-range strain effects in Pd NC's have been effectively quantified in the past via the Debye-Waller coefficient (DW)[29] and the Warren plots[30] respectively. These experimental results, along with the use of DFT and MD simulations with state-of-the-art reactive forcefields, allow us to present a complete qualitative account for the modification of the strain-state in Pd NC's as a result of $O_2$ adsorption.

## 2. Experimental and computational methodology

### 2. 1 Materials, synthesis and characterization

The Pd nanocubes were synthesized following a recipe extensively described in the literature[30,31]. The process was started by dissolving 105 mg of the capping agent polyvinylpyrrolidone (PVP), 600 mg of potassium bromide(KBr), and 60 mg of L-ascorbic acid in a ~20 ml glass scintillation vial in a total volume of 8.0 ml of deionized water. This was then equilibrated by stirring for around 10 minutes at a rate of 500 rpm inside an oil bath set at 373 K. A separate solution containing 57 mg sodium tetrachloropalladate (Na2PdCl4) dissolved in 3.0ml deionized water was then injected rapidly into the already prepared solution. The reaction was then allowed to continue for 3 hours, after which the particles were precipitated by adding a 10:1 acetone:water solution and centrifuging at 8000 rpm. The collected particles were then rinsed and similarly brought down twice more before being redispersed in 100 μl water as the final volume of solution used in the XRPD experiments. The samples were oxidized by leaving them exposed to ambient conditions.

The XRPD data was acquired at the powder diffraction beamline (11-BM) of the Advanced Photon Source at the Argonne National Laboratory. The capillary containing the Pd NC's were mounted on the beamline spinner and operated at 4200 rpm to ensure uniform collection of data and statistical consistency. Three measurements, 1 hour each, were performed at 100 K, 200 K and 300 K, with radiation of wavelength 0.0413874 nm, and the data was collected over a 2θ range of 0 to 60. The XRPD patterns thus obtained were then analyzed according to the WPPM method using the PM2K software (University of Trento, Italy, Scardi et al.[32] and references therein). Effects on the pattern due to the sample holder was modelled using pseudo-Voigt curves over the pattern from an empty Kapton capillary, while the instrumental profile was modelled using a standard $LaB_6$ powder pattern. The size and strain-broadening models were the same for all three temperatures. The initial size and shape of the NC was obtained from previously published TEM images[30], with a subsequent refinement of the mean and variance of the particle edge length (defined by a lognormal distribution), as well as the degree of truncation being performed by the WPPM algorithm. The strain broadening in particular is phenomenologically modelled with the WPPM approach, which accounts for the inhomogenous atomic displacement due to the surface of the NC's[30]. This effect, coupled with the shape and the truncation of the crystalline domain within the surfaces of the NC leads to anisotropic broadening of the diffraction peak[28]. The background was treated using a Chebyshev polynomial and the Pd unit cell parameter and Debye-Waller coefficients were independently refined to account for the effect of temperature, with the decrease in Bragg scattering intensity being accounted for with the simple scalar form $B_{iso}$ of the traditional Debye-Waller factor. Additionally, temperature diffuse scattering (TDS) is used in order to account for the local atomic displacement due to thermal (dynamic) as well as disorder (static) effects[33]. The phase analysis to separate the PdO phase from the metallic Pd was instead performed using the software Topas[32,34].

## 2.2 Classical molecular dynamics

The large-scale atomic/molecular massively parallel simulator (LAMMPS)[35] was used to perform classical MD simulations. To study the long-range strain we use Pd NC's modelled using embedded atom method (EAM) potentials[36–38]. It has been demonstrated in the literature[30] that these potentials produce a description of strain in reasonable overall qualitative agreement with experimental observations from XRPD, thus being quite adequate for modelling the long range strain generated due to a rounding of the Pd NC. The time step for the Verlet integration was taken as 1.5 femtoseconds (fs). The energy of the NC's was first minimized using a Hessian-free truncated Newton (HFTN) algorithm. The NC's were then thermalized by allowing them to evolve in a canonical ensemble (constant particle number N, system volume V and temperature T) until an appropriate thermal equilibrium was reached at T=300K. A chain of 100 Nosé-Hoover thermostats with a 5.0 picosecond damping parameter with MTK corrections[39] implemented on the equations of motion was used to reach the isothermal condition. Subsequently the constant temperature condition was removed, and the system was allowed to evolve in the microcanonical ensemble (constant N, V and total energy E). The simulation trajectories were visualized with Ovito[40] and the mean square displacement ($<u^2>$) of the system was then calculated using the software VMD[41].

On the other hand, to study the oxidation of Pd NC's and the short-range strain developed thereof, the ReaxFF forcefield developed by Senftle et al[14] was used. This forcefield describes coulombic, covalent and van der Waals interactions using bond-length/bond-order combined with polarizable charge[42]. The force field has been trained against ab initio data sets and can therefore accurately model bond breaking and formation in chemical reactions, and has been used in the past to model oxygen adsorption on Pd nanoparticles[43,44]. The Pd NC's of different size were first equilibrated at 300 K for 50 picoseconds (ps), and subsequently heated up to 600 K in another 50 ps using a chain of 100 Nosé-Hoover thermostats. Subsequently, in different simulations, they were either allowed to evolve at 600 K in a microcanonical (NVE) ensemble for 50 ps, or heated up again to 900 K in the same time. A significant difference in the ReaxFF-MD simulations compared to MD with EAM potentials, is the use of a much smaller integration timestep of 0.2 fs, since to smoothly model chemical reactions the timestep needs to be an order of magnitude lower than the smallest time-period for a molecular oscillation (~1 fs)[Hong, Hong 21].

## 2.3 Density functional theory

The initial step of the DFT calculations involved the relaxation of the geometry of the different Pd surface facets using the electronic structure calculation package Quantum Espresso[45,46]. The Perdew-Burke-Ernzerhof (PBE)[47] functional was used as an approximation for the exact electron-exchange correlation function, with the scalar-relativistic PBEsol pseudopotentials[48] using the projector augmented wave (PAW) method. These pseudopotentials have been reported in the literature[49–51] to faithfully reproduce elastic and mechanical properties in solids. Corrections for the van der Waals interaction were implemented using the Grimme-D2[52] method. A fine 4x4x1 Γ-centered Monkhorst-Pack k-mesh was used to sample the irreducible Brillouin zone, with a Marzari-Vanderbilt charge smearing of 0.001 Rydberg (Ry). The cutoffs used for the kinetic energy and charge density were respectively taken as 60 Ry and 480 Ry. The Pd 111, 110 and 100 surfaces were modelled using 3 x 3 surface slabs. The thickness of the surface slabs were 5 atomic planes in the Z-direction, with a 15 Å vacuum layer introduced to reduce interactions between periodic copies. The convergence thresholds for the energy and Hellman-Feynman forces were set to 0.001 eV and 0.01 eV/Å respectively. Charge and magnetization densities have been visualized using VESTA[53].

## 3. Experimental results

The analysis of the diffraction data using the WPPM approach can be seen in Fig 1 (a-c) at three different temperatures, with the insets showing the intensity in logarithmic scale in order to highlight the details in the peak tail regions. This serves as an indication of the quality of the fit, allowing us to draw reliable conclusions on the DW coefficient, in terms of $B_{iso}$ values. While the dynamic component of $B_{iso}$ dominates at higher temperatures, the static component due to distortion of the crystal lattice is a temperature-independent feature. The changes in the diffraction pattern due to the onset of oxidation can be clearly seen, most remarkably around the 15° mark on the Bragg angle scale in Fig 2(e).

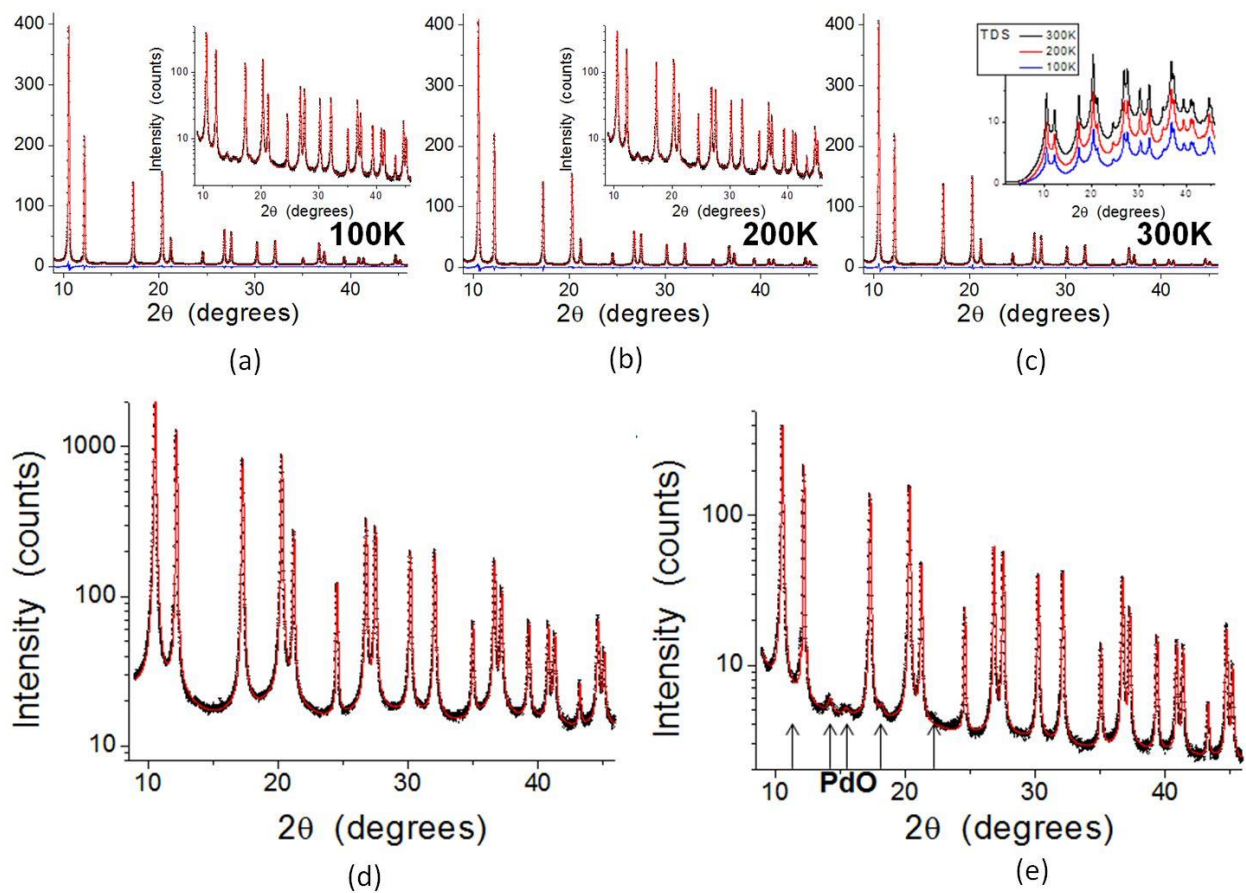

Figure 1: Whole powder pattern modelling of XRPD data: (a) and (b) and (c) show data (circle) and modelling (line) for the truncated cubes at 100 K, 200 K and 300 K respectively; the corresponding insets in (a) and (b) show the same data with the Y-axis (intensities) in the logarithmic scale for clarity, while the inset in (c) shows the TDS at the three temperatures; (e) and (f) show the fitting of the diffraction patterns (300 K, intensities in logarithmic scale) before and after the onset of oxidation respectively. Arrows in (e) point to the PdO peaks.

Using WPPM of the XRPD data, we study the change in morphology of the Pd NC due to oxidation. The size-broadening effect on the diffraction peaks from the NC's is modelled with a lognormal distribution for the NC edge size, with an additional freely refined parameter, going from 0 for perfect cubes to 1 for perfect octahedra, being used to quantify the deviation from a perfect cubic shape at the edges and corners of the NC. As is evident from Fig 2, oxidation of the NC leads to a reduction in average particle edge length from 14.23(1) nm to 14.05(1) nm, and an increase in degree of NC truncation from 0.110(2) to 0.163(4), suggesting an overall ~10% decrease in volume of the NC.

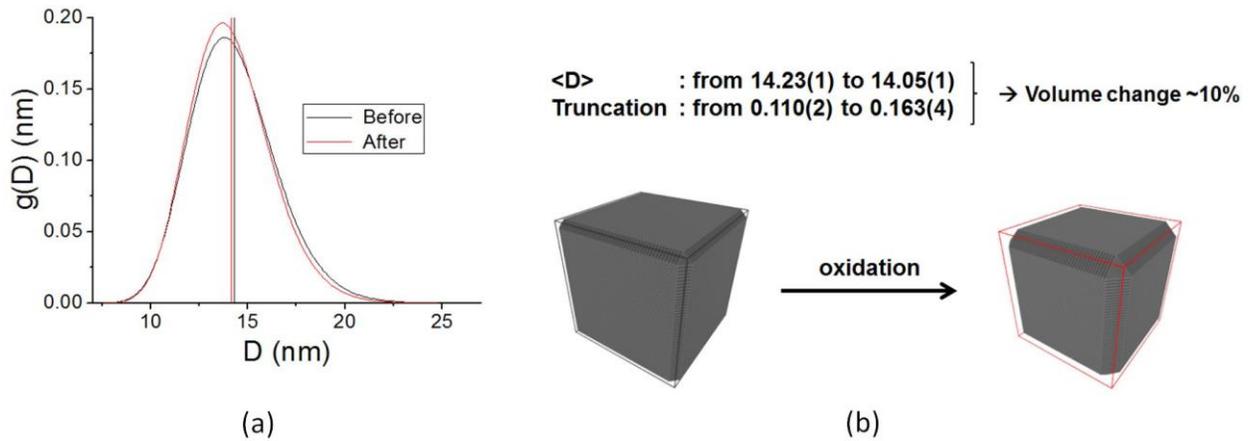

Figure 2: Change in NC morphology due to oxidation: (a) shows the lognormal distribution for NC edge size from WPPM modelling, while (b) is representative of the increase in truncation of the NC

This conclusion is further supported by performing a phase analysis of the diffraction pattern, which attests to the presence of a secondary nano-polycrystalline phase of around 7 % of palladium(II) oxide (PdO) formed on the Pd NC surface.

The short-range strain developed on the Pd NC surface, both static (deformation of the crystal structure) and dynamic (due to thermal vibrations of crystal atoms), is measured through the DW coefficient ($B_{iso}$). This is proportional to the mean-square displacement (MSD) of individual atoms in the NC from their ideal crystalline positions, and is connected to the MSD ($<u^2>$) through the expression, $B_{iso} = \frac{8\pi^2}{3}\langle u^2 \rangle$. It is thus a local property, mostly restricted to individual atoms, with a small but measurable effect on their closest coordination shells[54]. The $B_{iso}$ of nanocrystalline materials is known to be considerably higher than their bulk counterparts[55–57], and in a recent study Rebuffi et al (2020)[27] have demonstrated that its value for an NC is significantly increased by the surface disorder induced by adsorption of a capping agent.

From fitting the diffraction pattern, the values of $B_{iso}$ obtained for the Pd nanocube both before and after oxidation are observed to lie well above the bulk value for metallic Pd, as seen in Fig 3. This can be explained by a correlated Debye model[58], according to which the $B_{iso}$ at a particular temperature *T* is written as:

$$B_{iso}(T) = B_S + \frac{6h^2}{mk_B\Theta_D}\frac{1}{4} + \frac{6h^2}{mk_B\Theta_D}\left(\frac{T}{\Theta_D}\right)^2 \int_0^{\Theta_D/T} \frac{\xi}{e^\xi - 1}d\xi \quad [1]$$

where *h* is Planck's constant, $\Theta_D$ is the Debye temperature, m is the mass of an atom, the integral represents the (third) Debye function, and $B_s$ is an additional term which corresponds to static disorder.

While the quantitative description is somewhat restricted by the limited number of data points, it is nevertheless quite clear from the curves in Fig 3 that the $B_{iso}$ for the Pd nanocubes after oxidation is increased compared to the clean particle before. From these results, it is reasonable to make the assertion that it is in fact the adsorption of $O_2$ on the surface of the Pd nanocube which is responsible for the increase in $B_{iso}$, an assertion which we confirm using different simulation methods.

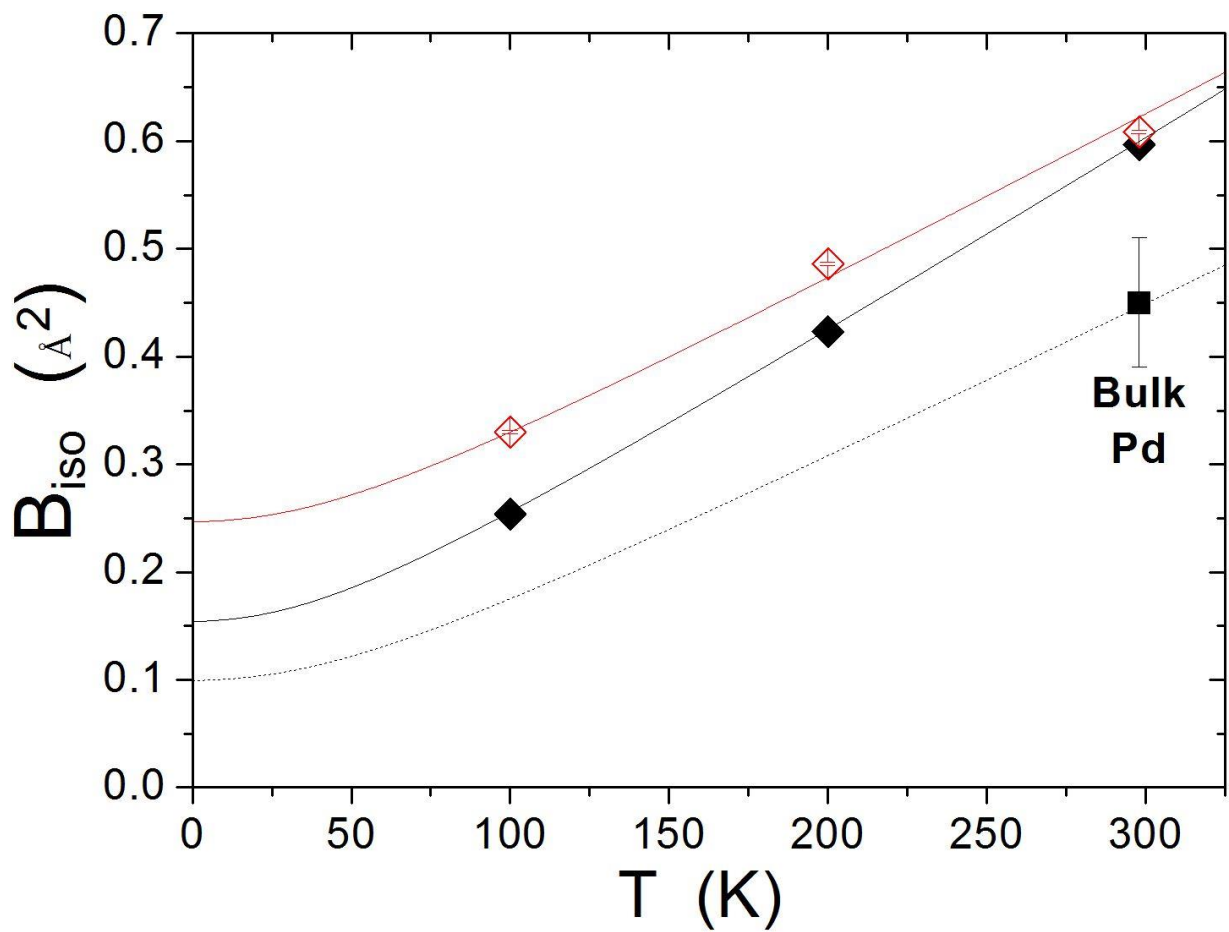

Figure 3: Increase in Debye–Waller coefficient ($B_{iso}$, left axis) with the measurement temperature. Solid black diamonds (◆) represent nanocubes before oxidation and open red diamonds (◊) after; the solid square (■) represents $B_{iso}$ of bulk Pd. Trends for $B_{iso}$ (dotted line) refer to a Debye model (see text for details)

Apart from the information on short-range strain due to oxidation, the long-range strain effects in the nanoparticle can also be quantified from fitting the diffraction pattern, and represented through the Warren plot, as seen in Fig 4(a). This diagram expresses the long-range strain within a crystalline domain by plotting the standard deviation of the distance L ($<\Delta L^2>^{1/2}$) between atomic scattering centers along specific crystallographic directions, as a function of L. In our specific case, it can be seen from Fig 4(a) that the ($<\Delta L^2>^{1/2}$) curve in the oxidized NC lies consistently lower compared to the clean particle, across all three main crystallographic directions [h00], [hh0], [hhh], suggesting a reduction in long-range strain in the NC due to oxidation.

The experimental results from Fig 3 and Fig 4(a) show that while oxidation leads to an increase in the short-range strain in the NC, as seen through the increase in the $B_{iso}$, the overall long-range strain represented by the inhomogenous atomic displacement in the Warren plot seems to diminish, compared to the clean particle. In the following sections we seek to further explain these seemingly contradictory changes in the strain-state of the Pd NC due to oxidation, through the application of multiscale modelling techniques.

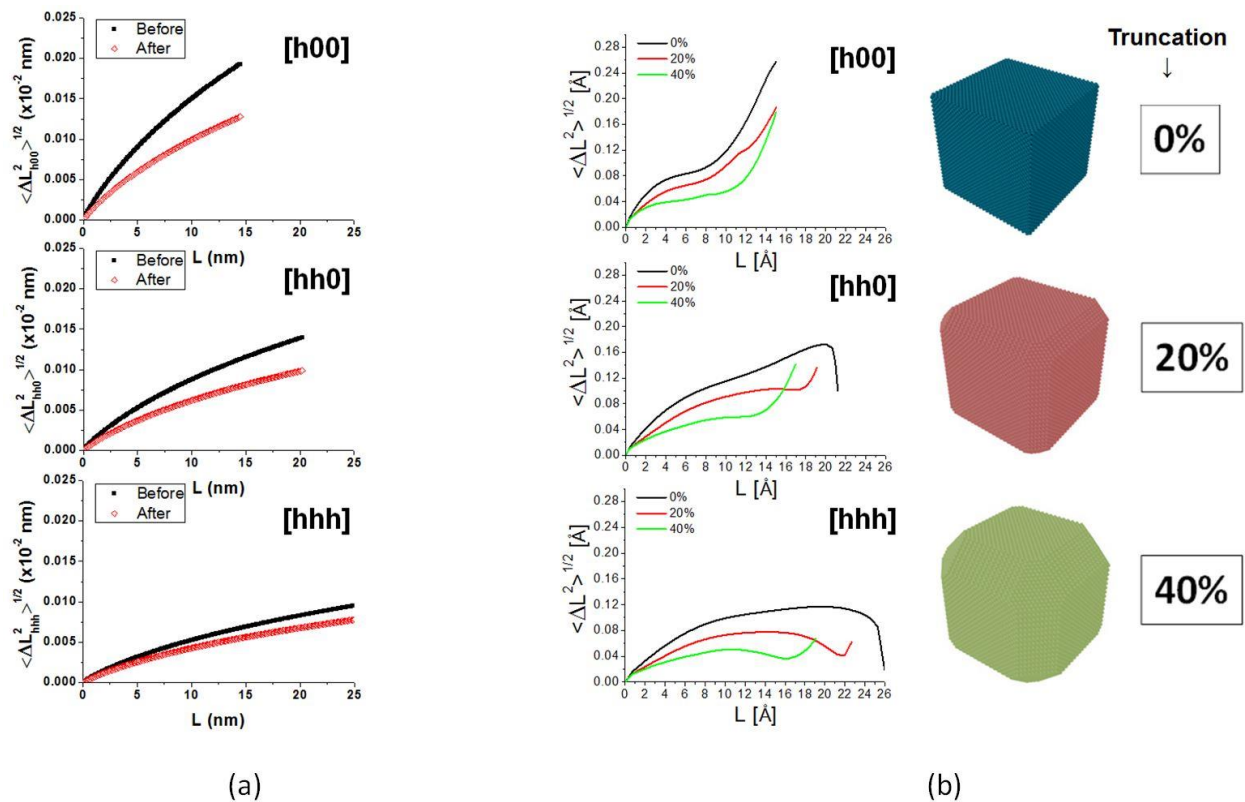

(a)            (b)

Figure 4: Warren plot describing long-range strain: (a) shows the experimental Warren plot, obtained from fitting of diffraction data before (black line) and after (green) oxidation, along different crystallographic directions and (b) shows the Warren plot obtained from MD simulations along the corresponding directions, with truncation of 0 (blue), 0.2 (red) and 0.4 (green).

**4. Progression of oxidation on Pd nanocube**

In order to study how the kinetics of the oxidation process correlate with the modification of the strain state in the Pd NC's, we have used MD simulations using reactive force fields (ReaxFF), which are significantly more computationally intensive than EAM potentials. In order to perform tractable simulations for the naturally slow oxidation process, we have made the simulations in an oxygen-rich environment, with a tradeoff of particle size in favour of longer simulation timescales. To investigate the kinetics of oxidation, we have used a Pd nanocube with 911 atoms. Figure 5 (also see supplementary video) shows the progressive oxidation of the NC. Concurrent with previous reports in the literature[21,24], we observe that the oxygen adsorption starts preferentially at edge and corner sites. Subsequently, these molecules are seen to split, and individual oxygen atoms can be seen to penetrate into the surface and subsurface layers of the Pd NC. As more and more oxygen penetrates the layers of Pd atoms, the cubic NC starts to lose its shape starting from the edges and corners, leading to the progressive 'rounding' of the NC shape, as observed experimentally by Collins et al[21], as well as our own diffraction results.

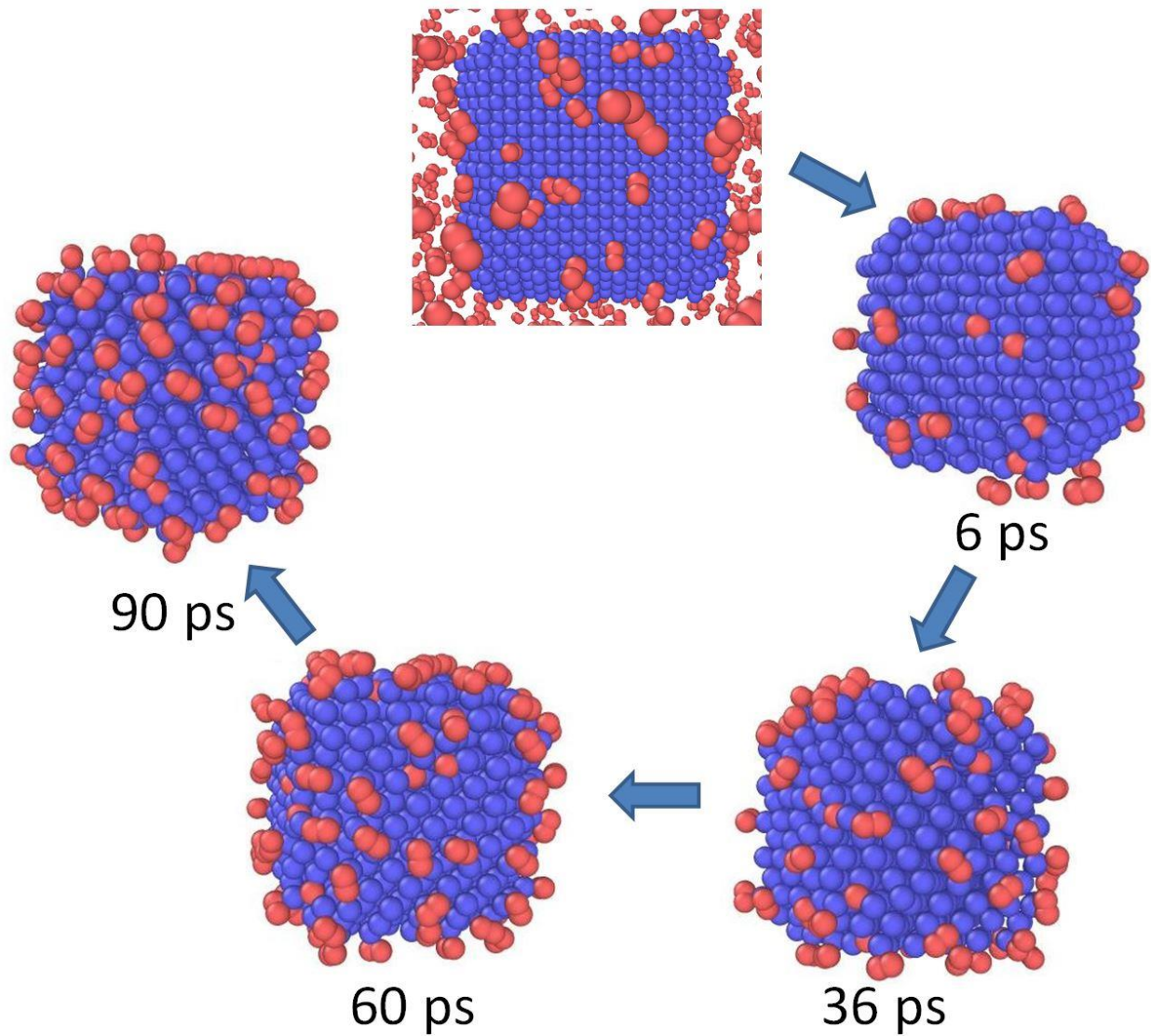

Figure 5: Oxidation on the Pd NC surface over time, starting from edge and corner sites. See supplementary video

To further characterize the oxidation process, we have calculated the radial pair distribution function (rpdf) of pairs of atomic species in the oxidized nanocube, after 20 ps, 430 ps and 780 ps. in the Pd-Pd rpdf in Fig 6(a), we notice a shift in the peaks towards higher distances as the simulation progresses, corresponding to a 'swelling' of the NC as increasing amounts of oxygen penetrate the surface and subsurface layer and form the PdO phase. This feature is particularly highlighted at longer distances, as seen in the inset of Fig 6(a). The presence and increase of the PdO phase is confirmed by the evolution of the O-Pd rpdf as shown in Fig 6(b). The position of this peak, centered close to 2.1 Å, attests to the formation of a partially crystalline phase of palladium oxide which continues to increase, as seen from the increase in peak height over longer simulation times. Fig 6(c) shows the O-O rpdf, with the two primary peaks showing the contraction and relaxation modes of the oxygen molecules in the atmosphere. The distribution of the O-O rpdf is distinctly asymmetric, with a 'step' in the upper tail region of the curve close to 1.3 Å. This step is representative of the oxygen molecules physisorbed on the surface of the Pd NC. It hints at a metastable state for oxygen on the Pd surface before complete dissociation, where the $O_2$ molecules remain weakly bound, with a higher interatomic separation compared to atmospheric $O_2$. The increase of the step height at longer time scales corresponds to the increase in the number of oxygen molecules physisorbed on the surface of the NC.

From a visual inspection of the MD trajectories combined with the calculated rpdf, we thus obtain a clear picture of how oxidation proceeds on a Pd NC – forming a layer of physisorbed oxygen on the surface with some molecules subsequently dissociating to form a PdO phase within surface and subsurface layers, with the process starting preferentially from edge and corner sites, thereby leading to a rounding-like modification of the NC shape.

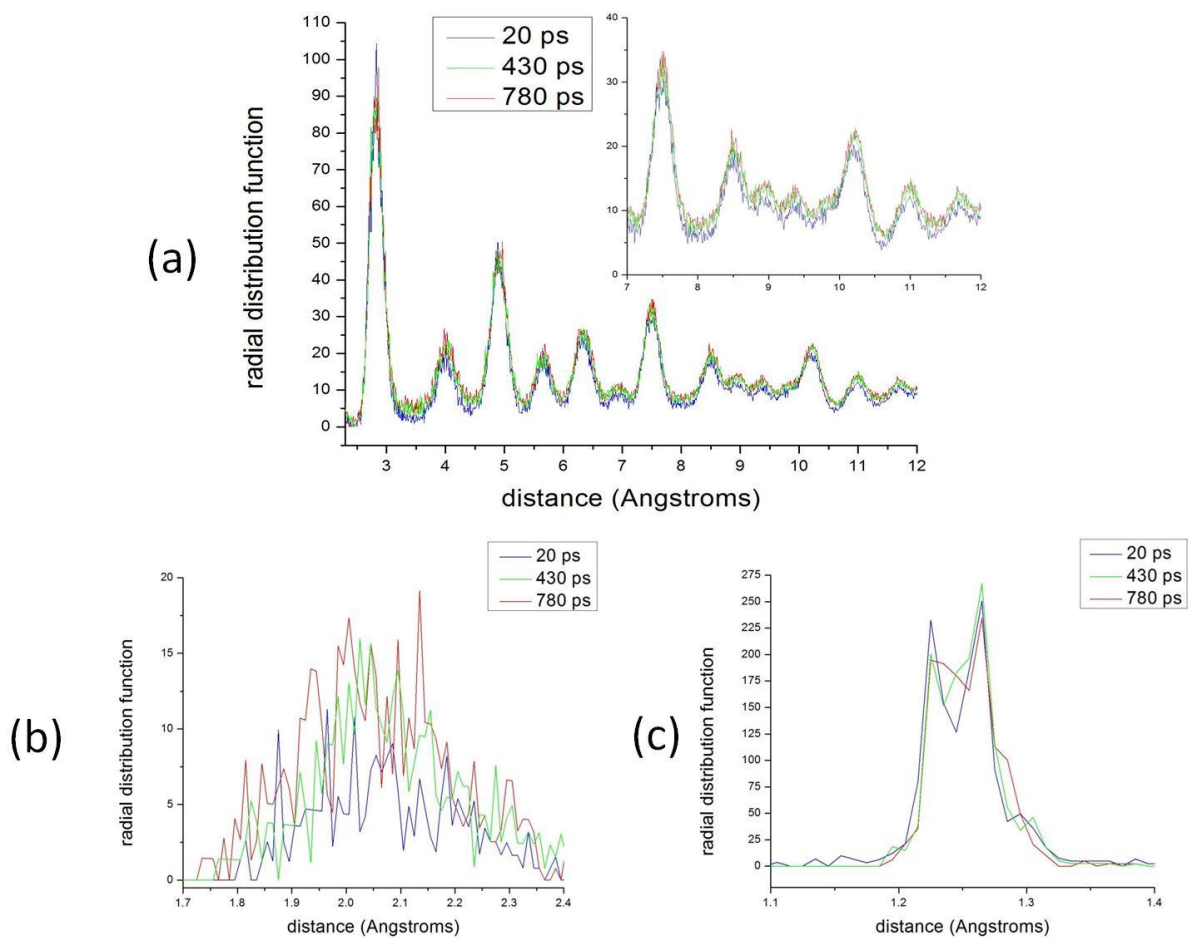

Figure 6. Radial pair distribution function: (a) denotes the Pd-Pd rpdf, with the inset highlighting the shifting of the peaks over longer distances as oxidation proceeds, (b) represents the O-Pd rpdf, showing the increase in the peak representing PdO at 2.1 Å with increasing oxidation, (c) shows the O-O rpdf, with a 'step' around 1.3 Å$^2$ denoting the physisorbed oxygen on the Pd surface. Blue, green and red lines correspond to the data at 20 ps, 430 ps and 780 ps of simulation time respectively.

In order to characterize the strain-state of the Pd NC, we have performed simulated oxidation on a larger Pd NC, with 11275 atoms. The increase in size of the simulated particle restricts the degree of oxidation occurring within the simulation timescale, but the larger particle allows for a more faithful representation of the elastic properties of the NC, suitable for a comparison with the significantly larger experimental objects.

Fig 7 shows the time evolution of atomic displacement magnitude on the surface of the Pd NC undergoing oxidation, given by the displacement of individual atoms from their mean position. As expected, high strain (corresponding to higher atomic displacement, red atoms) is generated at corner sites, due to preferential adsorption of oxygen at these locations and the under-coordination of atoms at these points. Over time, the high strain region expands to edge sites, before spreading to the rest of the NC surface, due primarily to undercoordination of the atoms at edge and corner sites coupled with the elastic anisotropy in Pd NC's[30].

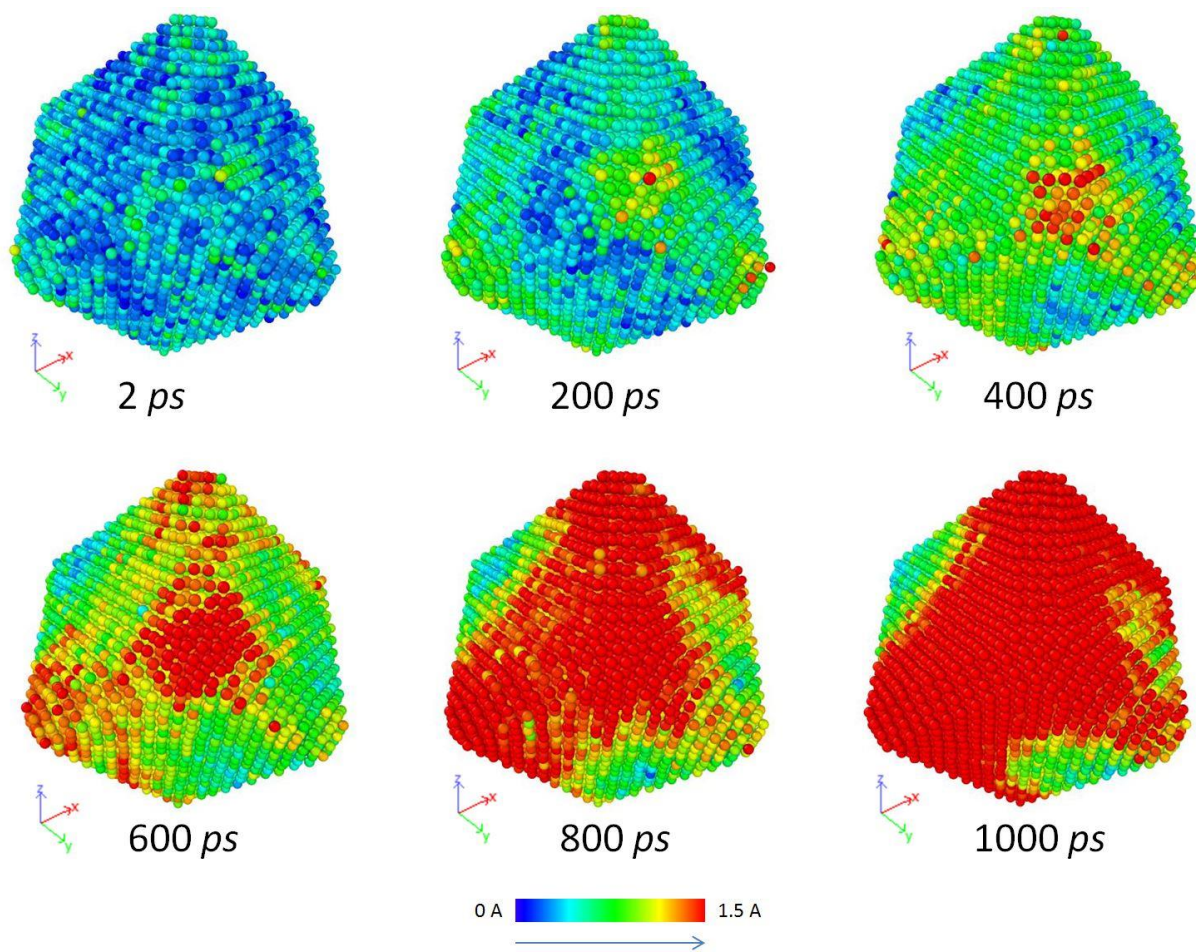

Figure 7. Increasing disorder on the Pd NC surface: The magnitude of atomic displacement due to oxygen adsorption, color-mapped to the surface of the Pd NC at six different points in simulation time.

This disorder introduced into the surface region of the NC due to the adsorption of oxygen and the resulting displacement of Pd atoms from their mean positions leads to the generation of short-range strain, which manifests itself in the static component of the Debye-Waller coefficient. This can be directly verified by obtaining the $B_{iso}$ via the mean square displacement (MSD) calculated from the MD trajectories.

Fig 8 shows a comparison of the $B_{iso}$ for the Pd NC from MD simulations under different conditions. For a clean NC evolving in a microcanonical (NVE) ensemble at 600 K (green line), the value of the $B_{iso}$ remains stable over the simulation time scale. In comparison, the $B_{iso}$ for the same NC evolving at the same temperature under the same NVE conditions in an oxygen atmosphere (blue line), is seen to steadily rise, corresponding to the increase in surface disorder as seen in Fig 7, as well as the experimental $B_{iso}$ curves obtained from XRPD (Fig 3). It is important to note here that the experimental conditions differ significantly from the simulation: the experiments were performed with larger particles oxidized over longer times, with measurements at lower temperatures compared to the MD calculations. All these factors contribute to a significantly higher value for the $B_{iso}$ obtained from the simulations. Nevertheless, the qualitative trend is clear – the presence of adsorbed oxygen on the Pd NC promotes surface disorder, leading to a clear rise in the Debye-Waller coefficient. In order to contrast the oxygen-induced static disorder with the dynamic disorder due to temperature, we show a third simulation, with the oxidation process of the same NC simulated on a temperature ramp from 600 K to 900 K (red line). The $B_{iso}$ in this case is seen to not only be much higher than at 600 K, but also appears to increase at a much faster rate, in contrast with the almost linear increase for the values at lower temperatures.

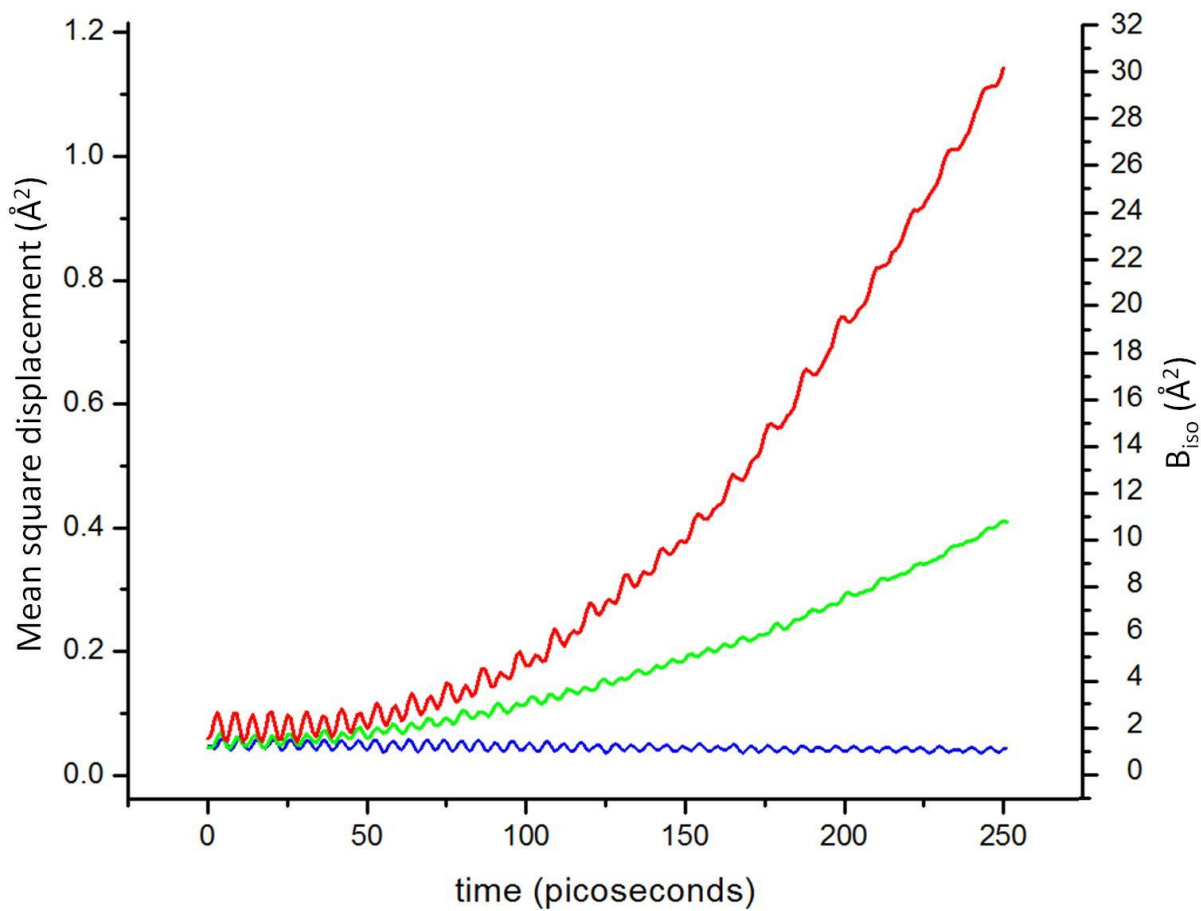

Figure 8: Time evolution of the Debye-Waller coefficient represented by $B_{iso}$ for a clean Pd nanocube at 600 K (green), a Pd nanocube undergoing oxidation at 600 K (blue), and a Pd nanocube undergoing oxidation while on a temperature ramp from 600 K to 900 K (red).

From the above discussion, it is then clear that oxidation in Pd NC's leads to an increased surface disorder as measured through $B_{iso}$, a distinct static effect clearly separate from the much stronger dynamic effects due to temperature. While the increase in $B_{iso}$ due to oxidation results from the displacement of Pd surface atoms from their mean position due to the interaction with oxygen atoms, the well-known temperature effect is due to the thermal motion of Pd atoms away from their mean positions, throughout the entire NC, causing the larger overall increase in $B_{iso}$.

As opposed to the increase in short-range strain, the long-range strain is experimentally observed to be lower in the oxidized NC compared to the clean particle as observed from the Warren diagram in Fig 4(a). To study the decrease in the long-range strain after oxidation, we use MD simulations using EAM potentials, which are more appropriate for simulating thermal/elastic properties.

To model the 'rounding-out' effect observed from experiment, we use three similar sized Pd nanocubes with increasing levels of truncation, going from 0 to 0.4. After each NC is thermalized, the Warren plot is calculated from the MD trajectories following the methods described in detail by Flor et al (2019)[57]. This process involves calculating the mean-square relative displacement (MSRD) from the averaged deviation in the distance between pairs of atoms $(\vec{r_{ij}})$ in the entire particle over the whole thermalization trajectory. This is then used to calculate the long-range strain leading to the Warren plots in Fig 4(b), which when compared with Fig 4(a), are clearly in a good qualitative agreement with experimental observations. We can then safely assert that the truncation of the nanocube due to preferential oxidation at edge and corner sites and the subsequent rounding in its shape, leads to a decrease in long-range strain. This can be explained as a release in strain due to a transition from a sharp to a more rounded morphology, reflected in the lower strain values for the oxidized/truncated particles in Fig 4 compared to the clean/sharp NC's.

## 5. Surface deformation with oxygen adsorption

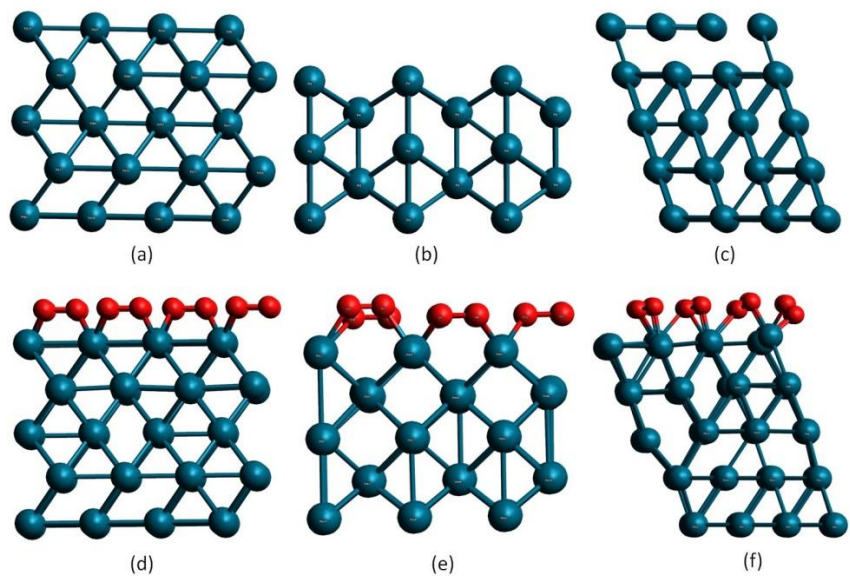

Figure 9. Deformation of the Pd nanocube surfaces in the presence of oxygen molecules: (a), (b) and (c) show the clean 100, 110 and 111 surfaces respectively; (d), (e) and (f) are the corresponding surfaces deformed in the presence of 1 ML $O_2$ surface coverage

In a recent study, Rebuffi et al[27] used DFT to show the deformation of a Pd NC surface region in the presence of a capping agent molecule, and demonstrated a corresponding increase in the Debye-Waller coefficient of the NC due to this surface softening. In a similar manner, we simulate the surface deformation of the Pd surface due to the adsorption of oxygen molecules in order to explain the higher $B_{iso}$ of the oxidized NC. This deformation is demonstrated by comparing the relaxed geometries, of the three main crystallographic facets 100, 110 and 111 of a Pd nanocube surface with 1 ML surface coverage (Fig 9), and calculating the MSD of the surface atoms with respect to the clean surface. The static component of the $B_{iso}$[57] is then calculated from the MSD, to provide a measure of surface-strain modification due to adsorption of oxygen.

The lightest deformation is observed on the low-density 100 surface (Fig 9a and 9d) which corresponds to the face of the Pd nanocube, with a calculated $B_{iso}$ of 0.318 Å². The edges of the nanocube are modelled with the 110 surface, and demonstrate a swelling effect (Fig 9b and 9e) upon oxygen adsorption, with an increased inter-layer separation between the atomic planes constituting the surface region, with a correspondingly high $B_{iso}$ of 4.73 Å². The 111 surface, corresponding to the corners of the truncated nanocube, shows the strongest deformation of all three surfaces. Deformation on this high-density surface takes the form of disorder in the atomic positions of the surface layers (Fig 9c and 9f), in response to the adsorption of $O_2$ molecules. The static $B_{iso}$ of the 111 surface, 5.67 Å², is the highest among all three.

The relatively weak deformation of the NC faces (100 surface), coupled with an outward expansion of the edges (110 surface) and strong disorder in the corner facets (111 surface), is congruent with the rounding-out of the cubic NC upon oxidation, as seen from the diffraction peak broadening, and reported in the literature from electron microscopy[21]. Additionally, to compare with our experimental results from XRD, we calculate an estimated $B_{iso}$ for the whole nanocube before and after oxidation, with the assumption that the $B_{iso}$ for the core of the NC corresponds to the bulk value (0.45 Å²). The total $B_{iso}$ of the NC can then be approximated as,

$$B_{iso,} \approx B_{iso,surface} \times f_{surf} + B_{iso,bulk} \times (1 - f_{surf})$$

where $f_{surf}$ is the fraction of total atoms estimated to be on the surface. It should be noted that this expression provides only a qualitative estimate of the $B_{iso}$ for a whole NC from small scale MD and DFT calculations; however, it has been previously used[29] to calculate the $B_{iso}$ for Pd NC's, with results reasonably comparable to experimental values.

For the clean nanocube, the main surface strain is due to dynamic (temperature related) effects, while after the adsorption of oxygen, the static effects can be considered to be dominant. As such, calculating the $B_{iso,\,surface}$ for the clean cube from MD and for the oxidized NC from DFT, we estimate $B_{iso,\,full}$ as 0.5312 Å² and 0.608 Å² respectively, compared to 0.597 Å² and 0.609 Å² from XRD at room temperature (Fig 3). While the model somewhat overestimates the difference in $B_{iso}$ between the clean and oxidized nanocubes, it is clear on a qualitative level that the surface softening and disorder induced by the adsorption of $O_2$ is ultimately responsible for the experimentally observed increase in $B_{iso}$.

Given the clear modification of the strain-state of the Pd nanocatalyst with oxidation as measured through X-ray diffraction, and the subsequent ascription of this effect to surface softening and disorder from $O_2$ adsorption, it is of some importance to understand how this surface disorder affects the reactivity of the adsorbed molecules.

Catalytic activity is well known[22] to vary on the different surface facets of Pd. Additionally, the variation in the calculated electronic density of states of $O_2$ molecules adsorbed on the Pd surfaces according to changing surface coverage (supplementary Fig S3), suggests that this is another factor which plays a role in modifying the catalytic activity. For a low surface coverage (0.25 ML) the average magnetization of $O_2$ molecules on the 100 surface, as calculated from DFT, is found to be less than half that on the 111 surface, in qualitative agreement with calculations by Long et al[22]. Interestingly however, we find that the average magnetization per molecule with surface coverage has opposite trends on the 100 surface, compared to the 111.

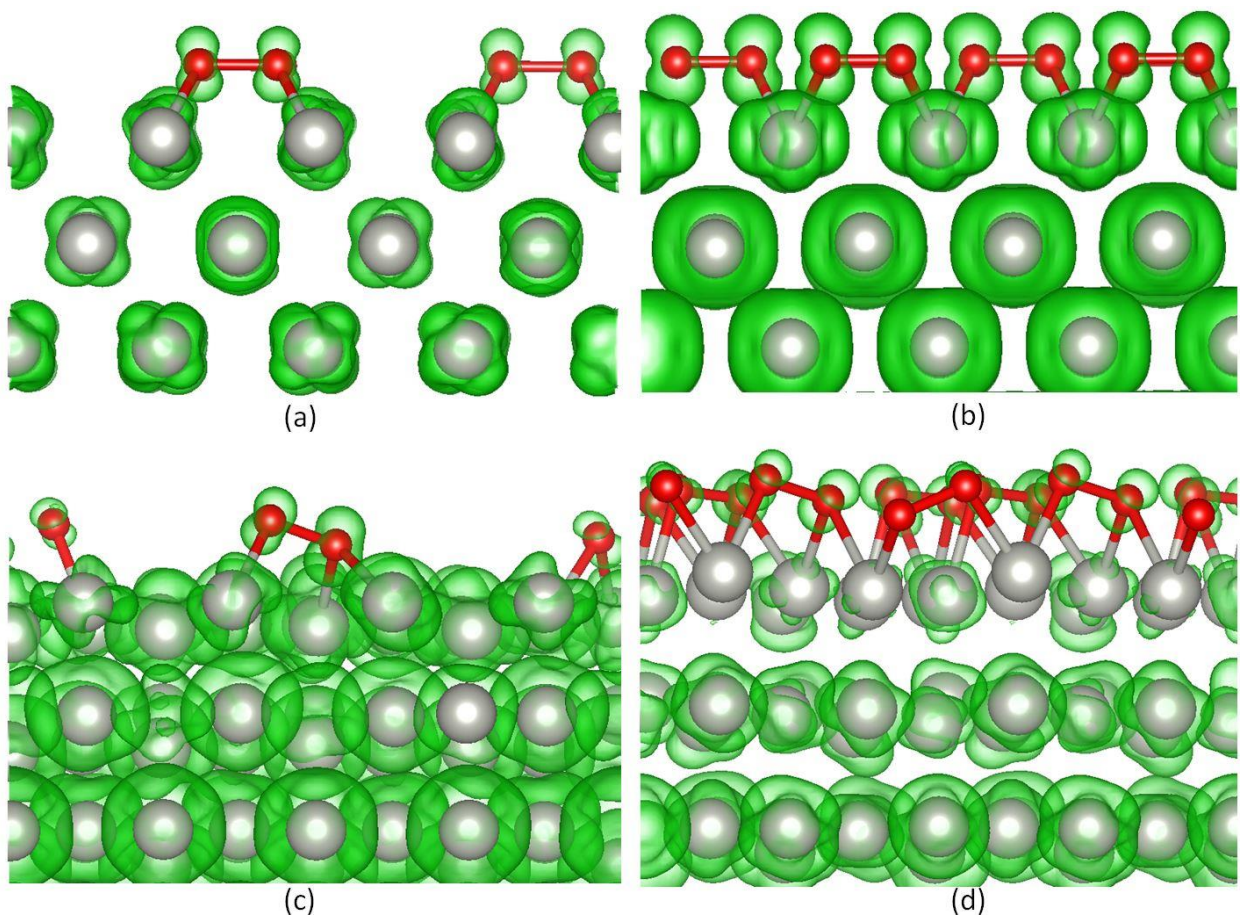

Figure 10. Magnetization density isosurface of O2 molecules on Pd surfaces: (a) and (b) show the density on the 100 surface at 0.25 ML and 1 ML surface coverage respectively, while (c) and (d) show the densities for the corresponding surface coverages on the 111 surface. Red balls correspond to oxygen atoms, grey balls palladium.

This effect can be seen clearly from the magnetization density (calculated as the difference in charge densities for spin states with opposite signs) in Fig 10. While the magnetization of $O_2$ molecules on the 100 surface is seen to go up with increasing surface coverage (Fig 10a and 10b), it appears to decrease for the molecules on the 111 surface (Fig 10c and 10d).

The magnetization of $O_2$ molecules is a critical measure of the catalytic activity of the Pd surface – transfer of charge from the metal surface to the half-filled $\pi^*$ orbitals of the paramagnetic ground state triplet $O_2$ molecule ($^3\Sigma_g$) leads to a reduction in the magnetic moment, and a corresponding increase in reactivity through a transition to the diamagnetic excited singlet states ($^1\Delta_g$, $^1\Sigma_g$). While at lower surface coverage, the 100 surface is predicted to perform better at activating O₂ molecules, it appears to become saturated with increased surface coverage. The 111 surface on the other hand seems to be able to reduce the magnetic moment of oxygen molecules at higher surface coverages.

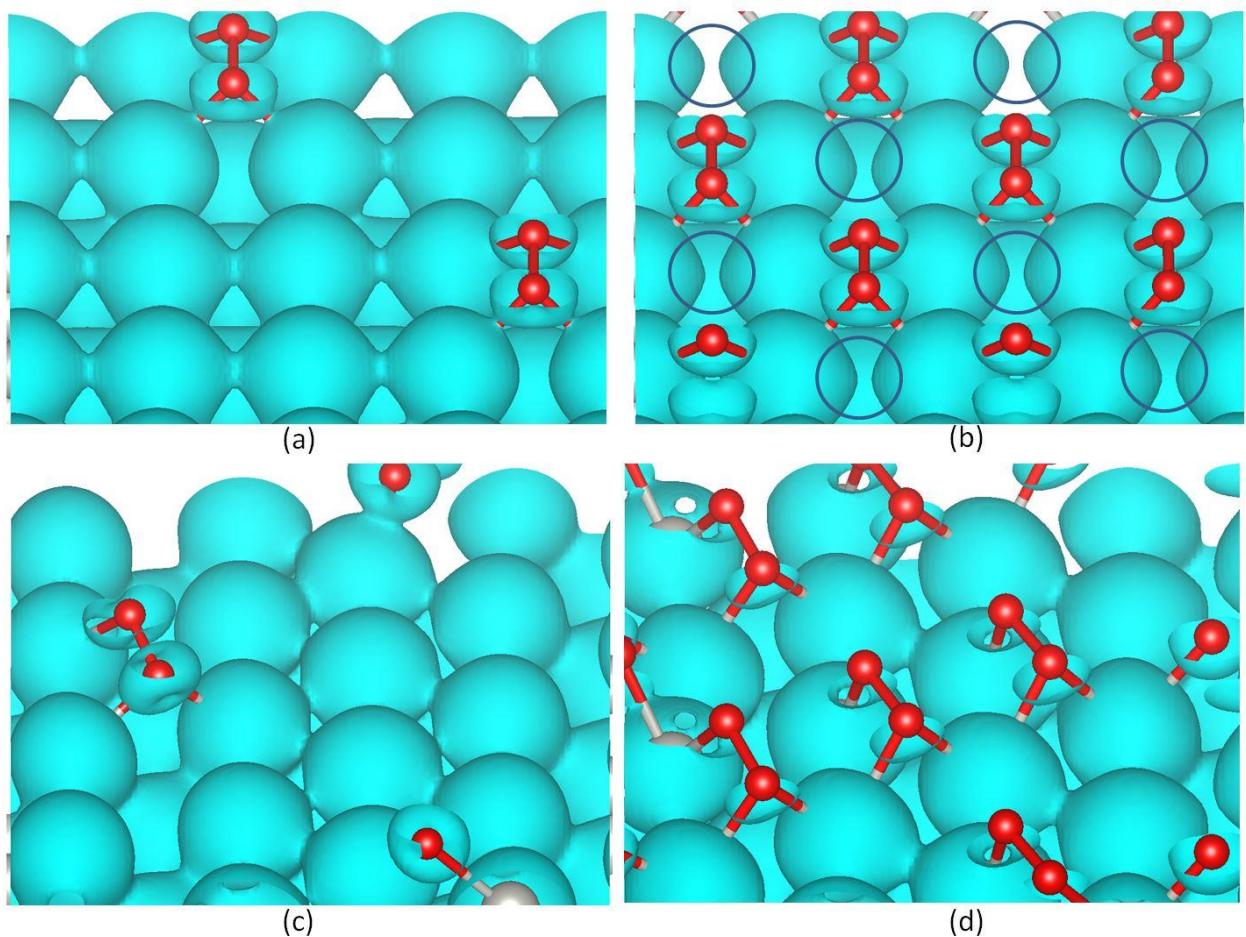

Figure 11. Partial charge density isosurface of Pd d-electrons: (a) and (b) show the density on the 100 surface at 0.25 ML and 1 ML surface coverage respectively, while (c) and (d) show the densities for the corresponding surface coverages on the 111 surface. Dark blue circles show the depletion in charge density. Red balls correspond to oxygen atoms, the isosurfaces are truncated above the Pd surface for clarity.

This effect can be explained by calculating the partial charge distribution of Pd d-electrons on the respective surfaces (Fig 11). On the 100 surface, we clearly observe a depletion of charge at higher surface coverage (Fig 11a and 11b), a distinct feature absent on the 111 surface. This depletion impedes the charge-transfer induced spin-flip process for the activation of O₂ molecules on the 100 surface at higher surface coverage. The absence of a similar depletion on the denser, disordered 111 surface, supporting a higher distribution of electrons, allows for a greater volume of charge transfer, explaining the decrease in magnetization of O₂ molecules at higher surface coverage. This phenomenon shows the potential for more sustained activation of O₂ molecules at corner sites of truncated Pd nanocubes, and Pd 111 surfaces in general.

## 6. Conclusion

Using a coupling of X-ray diffraction and theoretical modelling with classical MD and DFT, we study how the strain-state of cubic Pd nanocatalysts is modified as a result of oxidation. We show that the analysis of XRD data with sophisticated data analysis techniques like WPPM can be used as an accurate measure for strain effects in nanoparticles, both long-range through the Warren plot, and short-range via the Debye-Waller coefficient, even for a kinetic process such as oxidation. In this article, we find that the oxidation of Pd nanocubes leads to a decrease in long-range strain combined with a measurable increase of short-range disorder. Using classical MD simulations with EAM potentials, we show that the decrease in long-range strain is the result of the increasing truncation of the nanocube with oxidation. We analyse the kinetics of the oxidation process using MD simulations with ReaxFF force-fields, and show that the oxidation process and the accompanying local disorder progress from edge and corner sites of the NC, explaining the increased truncation of oxidized Pd nanocubes. From the MD simulations, we compute an increase in the Debye-Waller coefficient of the nanocube with progressing oxidation, in line with what is measured from XRD. Subsequently, we perform DFT calculations of oxygen adsorption on the individual crystallographic surface facets of truncated Pd nanocubes - 100 representing the faces, 110 the edges, and 111 the corners. We find that while the adsorption process has a limited deforming effect on the 100 faces, it leads to an outward expansion on the 110 edges and strong atomic disorder on the 111 surfaces. This combined deformation mechanism leads to a significantly large surface $B_{iso}$, which in turn leads to the increased $B_{iso}$ for the whole nanocube, in agreement with the XRD measurement, and provides a theoretical explanation for the experimentally observed rounding effect of oxidation. Finally, from the electronic structure calculation with DFT, we find that the different exposed facets have different trends in catalytic activity – while the ordered 100 faces are better at activating $O_2$ molecules from singlet to triplet states at lower surface coverage, the disordered 111 surfaces have the potential for more sustained $O_2$ activation at higher surface coverages. Thus, through a combination of multiscale modelling and experimental analyses, we provide a novel understanding of how oxidation modifies the strain-state in Pd NC's, and the potential effects it may have on their catalytic activity.

**Author contributions**

BM was primarily responsible for writing the article making the MD-ReaxFF and DFT simulations and calculating the Debye-Waller coefficient, and contributed to the analysis of the experimental data. AF was primarily responsible for making the MD-EAM simulations and calculating the Warren Plot from these simulations, and contributed to the analysis of the experimental data. PS was primarily responsible for conceiving and supervising the work, analysing the experimental data, and contributed to the collection of experimental data, and supervised the writing of the article.

**Acknowledgements**

The work carried out at Argonne National Laboratory was supported by the US Department of Energy, Office of Science, Office of Basic Energy Sciences, under contract DE-AC02-06CH11357. We acknowledge the CINECA award under the ISCRA initiative, for the availability of high performance computing resources and support in making DFT calculations. The authors would like to acknowledge Dr. Narges Ataollahi for discussions regarding applications of Pd nanoparticles in catalytic processes.